# Geometric Phase and Superconducting Flux Quantization

Walter A. Simmons & Sandip S. Pakvasa
Department of Physics and Astronomy
University of Hawaii at Manoa
Honolulu, Hi 96822

Abstract

In a ring of s-wave superconducting material the magnetic flux is quantized in units of $\Phi_0 = \frac{h}{2e}$. It is well known from the theory of Josephson junctions that if the ring is interrupted with a piece of d-wave material, then the flux is quantized in one-half of those units due to a additional phase shift of $\pi$. We reinterpret this phenomenon in terms of geometric phase.

We consider an idealized hetero-junction superconductor with pure s-wave and pure d-wave electron pairs. We find, for this idealized configuration, that the phase shift of $\pi$ follows from the discontinuity in the geometric phase and is thus a fundamental consequence of quantum mechanics.

Geometric phase has been contained in quantum mechanics since the foundations of the field were set down in the early twentieth century; however, the phase and its importance were not recognized for some time. Pancharatnam[1] discovered the classical geometric phase in optics in 1956 and Berry's important 1987 quantum mechanics paper[2] stimulated the rapid development of the field. By 1992, Anandan[3], in a review article in *Nature*, was able to conclude that the phase had been convincingly demonstrated. The first application of geometric phase to Josephson Junctions was carried out by Anandan and Pati in 1997. They showed that the zero voltage tunneling supercurrent is geometric in nature and that it is proportional to the speed of the state vector in projective Hilbert space.[4]

The appearance of a phase discontinuity of $\pm\pi$ arising from geometric phase, under certain circumstances, was shown[5] to be a general feature of quantum mechanics in 2003, but has so far found only limited application[6]. Here we show that a well known phenomenon[7,8] in superconductivity, the quantization of magnetic flux in one half of the usual unit[9], which is $\Phi_0 = \dfrac{h}{2e}$, can be interpreted as an effect of the discontinuity in geometric phase. This phase shift in superconductors has been understood in terms of the physics of the Josephson junctions and the result has been applied to high temperature superconductors[10] in order to test the idea that they involve d-wave electron pairs. Our application considers an idealized hetero-junction superconductor with pure s-wave and pure d-wave electron pairs. We find, for this idealized configuration, that the phase shift of $\pi$ follows from the discontinuity in the geometric phase[4] and is thus a fundamental consequence of quantum mechanics.

Applications of quantum geometric phase have been made in nearly every branch of physics, from fundamental material science[11,12] to quantum computing with superconducting nanocircuits[13], as well as in chemistry[14], and it has been suggested that phase may become important in biology[15].

Since the phase has been long present, but not fully recognized, some applications entail reinterpreting known phenomena in terms of the phase. An important and illustrative example is the reinterpretation of the so called Guoy effect in optics as a geometrical phase[16], which we will summarize below.

An idealized superconducting ring, which consists of a composite of s-wave material and d-wave material, will, in the absence of external electromagnetic fields, exhibit quantization of the magnetic flux in units of one half of the usual unit, $\Phi_0 = \frac{h}{2e}$. This half-unit quantization of the magnetic flux will occur whenever there is an odd number of phase shifts of magnitude $\pi$ in the circuit. The theoretical argument[6] for the $\pi$ phase shift in a composite ring was based upon the dynamics of the Josephson junction and on thermodynamics, and has experimental support.

We next explain the quantum mechanics of the phase shift and then we shall proceed to reinterpret the s-wave/d-wave superconductor hetro-junction.

It has long been known[17] that when a light beam converges to a focus, then diverges again, the light experiences a phase shift whose magnitude depends upon the details. For example, for a beam with a Gaussian profile and a very small waistline at the focus, an abrupt phase shift of $\frac{\pi}{2}$ occurs for each of the two transverse directions, for a total phase shift of $\pi$. This result, which follows from standard classical electrodynamics, has been reinterpreted in terms of geometric phase[15]. For one transverse direction, the complex curvature of the wave reverses at the focus; the geometric phase, which is directly related to the curvature, changes by $\frac{\pi}{2}$.

That optical example of geometric phase is closely analogous to a well-known[18] phase flip of $\pm\pi$, which occurs in optics when the polarization of a light beam is rotated from some initial state, through and beyond, another polarization state that is orthogonal to the initial state.

The latter example of the geometric phase discontinuity has been shown to occur rather generally in quantum mechanics[4]. This can be understood by considering the behavior of a complex quantum state vector as it is impelled through a series of states in Hilbert space by an external force. Suppose the initial state is $|\Psi_i\rangle$, the final state is $|\Psi_f\rangle$, and some intermediate state $|\Psi_0\rangle$ is orthogonal to the initial state, $\langle\Psi_0\|\Psi_i\rangle=0$. In the complex plane, the trajectory is a sequence of projections of each state upon its subsequent state. The trajectory from the initial to the final state passes through the origin (with a positive or negative infinitesimal imaginary part); the phase goes through an inverse-tangent singularity and changes by $\pm\pi$.

Finally, turning to superconductivity, we adopt the theoretical framework[19], in which superconductivity is viewed as a consequence of the breaking of gauge invariance entailing the formation of Cooper pairs. We consider a ring of material in which the supercurrent is carried by s-wave Cooper pairs. The ring is interrupted by a section of material in which d-wave Cooper pairs carry the supercurrent. The supercurrent passing through the idealized hetro-junction experiences a shift of $\pm\pi$ in the geometric phase due to the orthogonality of d-wave and s-wave. Since the orientation of the d-wave relative to the s-wave is not meaningful, we have no sum over dimensions as in the optical beam analogy; a shift of $\pm\pi$ is the result of a single inserted section of material and the half-unit magnetic flux quantization follows.

Since the 1997 work of Anandan and Pati, it has been known that the zero-voltage current in a tunneling supercurrent arises from the geometry of Hilbert space and is independent of the specific Hamiltonian, (which is a general feature of geometric phase[20]).

Recently, experiments on hetero-junctions have supported the idea that high $T_c$ materials are d-wave superconductors. While we are not discussing here realistic models of high $T_c$ materials here, our results show that a phase shift of $\pm\pi$ in s-wave/d-wave hetero-junctions arises from the fundamentals of quantum mechanics.

---

[1] Pancharatnam, S., *The Proceedings of the Indian Academy of Sciences*, Vol XLIV, No. 5, Sec. A, 247 (1956) in *Collected Works of S. Pancharatnam*, Oxford University Press, London (175).